# Improving Pediatric Low-Grade Neuroepithelial Tumors Molecular Subtype Identification Using a Novel AUROC Loss Function for Convolutional Neural Networks

Khashayar Namdar[1,2,4,10], Matthias W. Wagner[1,2,5], Cynthia Hawkins[6], Uri Tabori[7], Birgit B. Ertl-Wagner[1,2,3,4], Farzad Khalvati[1,2,3,4,8,9,10]

Affiliations:

1.  Department of Diagnostic Imaging & Image-Guided Therapy, The Hospital for Sick Children (SickKids), Toronto, ON, Canada

2.  Neurosciences & Mental Health Research Program, SickKids Research Institute, Toronto, ON, Canada

3.  Department of Medical Imaging, University of Toronto, Toronto, ON, Canada

4.  Institute of Medical Science, University of Toronto, Toronto, ON, Canada

5.  Department of Diagnostic and Interventional Neuroradiology, University Hospital Augsburg, Germany

6.  Department of Paediatric Laboratory Medicine, Division of Pathology, The Hospital for Sick Children, University of Toronto, Canada

7.  Department of Neurooncology, The Hospital for Sick Children, Toronto, ON, Canada

8.  Department of Computer Science, University of Toronto, Toronto, ON, Canada

9.  Department of Mechanical and Industrial Engineering, University of Toronto, Toronto, ON, Canada

10. Vector Institute, Toronto, ON, Canada

Corresponding Author: Farzad Khalvati, email: farzad.khalvati@utoronto.ca

*Abstract* **Pediatric Low-Grade Neuroepithelial Tumors (PLGNT) are the most common pediatric cancer type, accounting for 40% of brain tumors in children, and identifying PLGNT molecular subtype is crucial for treatment planning. However, the gold standard to determine the PLGNT subtype is biopsy, which can be impractical or dangerous for patients. This research improves the performance of Convolutional Neural Networks (CNNs) in classifying PLGNT subtypes through MRI scans by introducing a loss function that specifically improves the model's Area Under the Receiver Operating Characteristic (ROC) Curve (AUROC), offering a non-invasive diagnostic alternative. In this study, a retrospective dataset of 339 children with PLGNT (143 BRAF fusion, 71 with BRAF V600E mutation, and 125 non-BRAF) was curated. We employed a CNN model with Monte Carlo random data splitting. The baseline model was trained using binary cross entropy (BCE), and achieved an AUROC of 86.11% for differentiating BRAF fusion and BRAF V600E mutations, which was improved to 87.71% using our proposed AUROC loss function (p-value 0.045). With multiclass classification, the AUROC improved from 74.42% to 76. 59% (p-value 0.0016).**

*Keywords pediatric low-grade glioma, brain tumor, loss function, machine learning*

## I. Introduction

Pediatric Low-Grade Neuroepithelial Tumors (PLGNT) are the most common type of brain tumor in children, comprising about 40% of brain tumors in this age group [1], [2], [3]. The prognosis of PLGNT varies widely based on their molecular subtypes, making the identification of BRAF status crucial in treatment planning, especially with the emergence of BRAF-specific therapies [4].

Around 20% of PLGNT patients are affected by the BRAF-V600E mutation, which is linked to lower survival rates, particularly when accompanied by CDKN2A deletion [5]. In contrast, tumors with BRAF fusion generally have a better prognosis, leading to a lower-risk classification for those patients. Therefore, differentiating between BRAF and non-BRAF tumors is essential for effective treatment and risk stratification in PLGNT, particularly with the introduction of new BRAF-targeted therapies [6].

The current standard for identifying PLGNT subtypes is through biopsy or surgical resection. However, these methods are invasive and come with several complications, such as infection or hemorrhage [7], [8], [9], [10]. Additionally, a biopsy may not always be possible depending on the tumor's location.

Medical imaging offers a non-invasive alternative, capable of visualizing the entire tumor, but it has limitations in accuracy for certain tumor types and radiologists cannot determine molecular subtypes on their own [11]. Therefore, reliable and precise imaging-based molecular subtyping could be invaluable in accurately selecting patients for BRAF-targeted treatments and clinical trials [12], [13].

Recently, Machine learning (ML) has taken a leading role in medical image analysis, successfully applied in tumor segmentation, outcome prediction, and tumor subtype identification. In this study, a Convolutional Neural Network (CNN) is trained to differentiate BRAF mutated, BRAF fused, and non-BRAF tumors using MR images. Using novel loss functions, the performance of the baseline CNN is improved in binary and multiclass classification contexts and the loss functions are published as a Python library.

## II. RELATED WORKS

### A. Non-invasive PLGNT Subtype Identification

There is limited prior research on classifying PLGNT using ML, but several notable clinical studies have been conducted. Among them, Wagner et al. employed radiomics to differentiate pretherapeutically between BRAF-mutated and BRAF-fused tumors [14]. In their study, they analyzed T2 Fluid-Attenuated Inversion Recovery (FLAIR) MR images from 115 pediatric patients across two institutions. A pediatric neuroradiologist annotated the regions of interest, from which radiomics features were extracted. The study used a random forest (RF) algorithm as the binary classifier. To evaluate their models, they used 4-fold cross-validation on 94 patients from one hospital and a separate external validation on 21 patients from the other hospital. The Receiver Operating Characteristic (ROC) curves from these models indicated an average internal area under the ROC curve (AUROC) of 0.75 and an external AUROC of 0.85. Their findings highlighted that both tumor location and the patient's age were important indicators of BRAF status in the first cohort. Incorporating age and location with radiomics features led to an improvement in the average AUROC to 0.77.

In a different study, Wagner et al. examined a more extensive collection of 251 PLGNT FLAIR MR images [15]. Their focus was on how data splits and the initialization of ML classifiers affect performance, specifically looking at the average AUROC and its variability. Remarkably, they achieved comparable outcomes using only 60% of the training data, with an average AUROC of 0.83, which was close to the 0.85 AUROC achieved by models using the entire dataset.

Xu et al. conducted a study on a dataset of 113 PLGNT patients, including 43 with BRAF V600E mutations and 70 with other subtypes, using radiomics to identify BRAF mutations [16]. They divided the dataset into a training and testing group (70/30 split) and employed a 5-fold cross-validation to refine their pipeline within the training set. RF was found to be the most effective classifier. The study reported an average training AUROC of 0.72, with a 95% confidence interval (CI) of 0.602 to 0.831, and a testing AUROC of 0.875. They found tumor location to be a crucial indicator of BRAF mutation, with the addition of radiomics improving the average training AUROC to 0.754 (95% CI of 0.645 to 0.844) and the testing AUROC to 0.934. Interestingly, the side of the tumor location (left vs. right) did not significantly predict BRAF mutation.

Kudus et al. utilized a dataset comprising 253 PLGNT patients and introduced a thresholding approach to enhance the reliability of ML binary classifiers differentiating between binary BRAF-mutated and BRAF-fused tumors [17]. By incorporating radiomics features from MR images and clinical data such as sex, age, and tumor location, they achieved a 92.2% accuracy rate in 80.7% of patients classified with high confidence. Their research notably reached a peak AUROC of 0.925.

Namdar et al. used CNN models to identify PLGNT BRAF subtypes [18]. In a retrospective study, they analyzed MRI FLAIR sequences from 143 BRAF fused and 71 BRAF V600E mutated tumors. The authors created 3D binary Volumes of Interest (VOI) masks for each class to establish probability density functions (PDFs) of tumor location. Three distinct pipelines were developed and evaluated: one based solely on location, another on CNN, and a hybrid approach. The findings revealed that the location-based classifier attained an AUROC of 77.90 (95% CI: 76.76, 79.03). The CNN-based classifiers yielded an AUC of 86.11 (CI: 84.96, 87.25). Notably, the hybrid models that combined tumor location guidance with CNNs surpassed these results, achieving an average AUROC of 88.64 (CI: 87.57, 89.72), a difference confirmed as statistically significant (Student's t-test p-value 0.0018).

### B. AUROC for Optimizing and Evaluating ML Models

AUROC is employed in two principal capacities. It is widely used as a metric for assessing the efficacy of binary classifiers, especially in the field of medicine and with imbalanced datasets. Additionally, AUROC can be applied as an objective function, or loss function, which is instrumental in the optimization and training of predictive models [19]. However, AUROC is not differentiable, and violates requirements of Gradient Descent, the dominant approach for optimizing ML models during training [20]. Thus, the challenge is to propose a diffentiable AUROC or utilize other frameworks, such as Genetic Algorithm (GA) that allow direct optimization of models for AUROC [21].

Namdar et al. introduced a variant of AUROC that incorporated model confidence into account [19]. They demonstrated the modified AUROC was linked to Binary Cross Entropy (BCE) loss function and would be a better metric for monitoring CNNs during training. The approach was validated using three datasets and was shown to be effective.

Yan et al. highlighted the limitations of conventional loss functions such as cross entropy (CE) and mean squared error (MSE), which are traditionally used to enhance correct classification rates but might not efficiently optimize models when the aim is to boost the classifier's ability to distinguish between different outcomes across various decision thresholds [22]. The study introduced an objective function, derived as a differentiable approximation to the Wilcoxon-Mann-Whitney (WMW) statistic, which aligns directly with the AUROC.

Gajić et al. proposed a differentiable variation of AUROC that allows optimizing models directly for optimal AUROC

performance [23]. Their approach is based on two techniques: integral to series (Riemann sum) for calculating area under the curve and using sigmoid to avoid cusps.

## III. METHODOLOGY

In this research, a shallow 3D CNN architecture was utilized to differentiate BRAF fused and V600E mutated tumors using VOIs at patient level. Appending 125 non-BRAF patients to the dataset, enabled conducting multiclass PLGNT subtype identification. To derive a differentiable AUROC loss function for binary and multiclass classification, this research is inspired by the WMW statistic and utilizing sigmoid to eliminate cusps in AUROC calculations.

### A. Dataset

Our REB-approved retrospectively acquired study cohort consisted of 143 BRAF fusion, 71 BRAF V600E mutation, and 125 non-BRAF children with PLGNT. As the input to the CNNs, we used 3D MRI Fluid-Attenuated Inversion Recovery (FLAIR) sequence and the tumor segmentations provided by a pediatric neuroradiology fellow and verified by a senior pediatric neuroradiologist.

### B. Data Split and Hyperparameters

To measure the randomness of the AUROC measurements, we used a Monte Carlo data splitting, as shown in Figure 1. This procedure involved random splitting of the dataset into a train/validation/test set with a 60/20/20 ratio, and was repeated 100 times. The design of the architecture and hyperparameter tuning are elaborated elsewhere [18]. The optimal hyperparameters for the baseline and the proposed pipeline are outlined in Table 1. For the baseline, the batch size was 8 and CE was used as the loss function. For the proposed pipeline the batch size was increased to 64.

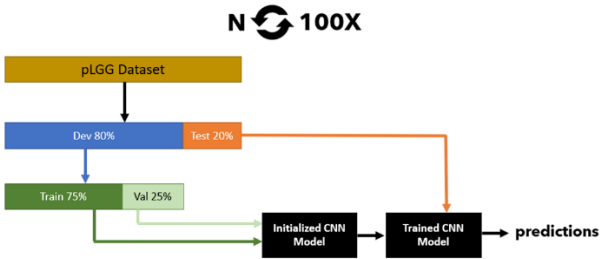

Fig. 1. The Monte Carlo data spliting approach

TABLE I. GRID SPACE OF HYPERPARAMETERS

| Hyperparameter | Value |
| --- | --- |
| maximum number of epochs | 40 |
| learning rate | 0.1 |
| optimizer | stochastic gradient descent (SGD) |

### C. Development Environment

The models were developed using PyTorch version 1.10.2 within a Python 3.9.7 setting, supported by cuda 11.3. The computational hardware consisted of two GeForce RTX 3090 Ti GPUs on a Lambda Vector GPU workstation.

### D. Proposed Loss Function

Deriving ROC curve and calculating the area under the curve is a classic, but inefficient method of estimating AUROC. AUROC is a non-parametric metric that indicates goodness of ranking, and WMW statistics, another non-parametric metric for ranking [22], provides the context for defining AUROC independent of ROC. The key is the fact that AUROC is a probability. Assuming one random positive case and one random negative case are chosen, AUROC is the probability of observing a higher prediction score for the positive case.

ALGORITHM I. PSEUDOCODE OF DEFINING AUROC AS A PROBABILITY

```
actual_positives_pred  # list of predicted probabilities for actual positives
actual_negatives_pred  # list of predicted probabilities for actual negatives
AP = len(actual_positives_pred)  # Number of actual positives
AN = len(actual_negatives_pred)  # Number of actual negatives
AUC = 0

# Iterate over all pairs of actual positive and negative cases
for positive_case_pr in actual_positives_pred:
    for negative_case_pr in actual_negatives_pred:
        # Compare predictions for a positive and a negative case
        if positive_case_pr > negative_case_pr:
            AUC += 1
        elif positive_case_pr == negative_case_pr:
            AUC += 0.5
        else:
            AUC += 0

AUC = AUC / (AP * AN) # Scale the metric to the range [0, 1]
```

Algorithm 1 is an if-else-based implementation of AUROC, which is neither efficient nor helpful in highlighting why AUROC is not differentiable. Algorithm 2 is our improved implementation of AUROC in PyTorch. Unit_step in Algorithm 2 is the unit step function, also known as Heaviside step function, which is defined as (1).

ALGORITHM II. EFFICIENT AUROC IMPLEMENTATION IN PYTORCH

```
def AUC_metric(y_score, y_true):
    pos_pred = y_score[y_true == 1]
    neg_pred = y_score[y_true == 0]
    pairwise_matrix = pos_pred.unsqueeze(1) - neg_pred
    return unit_step(pairwise_matrix).mean()
```

$$H(x) = \begin{cases} 1, & if \ x > 0 \\ 0.5, & if \ x = 0 \\ 0, & if \ x < 0 \end{cases} \quad (1)$$

Equation (1) uncovers why AUROC is not differentiable: The unit step function is not differentiable at x = 0. Thus, approximating the unit step with a differentiable function results in an AUROC loss function. Fourier Series, tanh, and arctan are among the options for approximating the unit step function. However, we chose logistic function, as elaborated in

(2), given its accuracy and computational efficiency. In (2), k is the logistic growth rate, and the higher it is, the better the function approximates unit step function. In this study we set α to 20. L is the supremum and $x_0$ determines the midpoint of the function. In this study, these values are set to 1 and 0, respectively.

$$f(x) = \frac{L}{1+e^{-k(x-x_0)}} \quad (2)$$

Algorithm III. shows our PyTorch implementation of AUROC loss function. It should be highlighted that we have incorporated Softmax function into the loss function, which aligns with PyTorch native approach for implementing loss functions. Also, given that Algorithm III. is a loss function, complement of AUROC is calculated as the loss. Algorithm IV. is an extension of Algorithm III. which is used for multiclass classification. The loss function is available in GenuineAI python library (https://pypi.org/project/GenuineAI/)

ALGORITHM III. IMPLEMENTATION OF AUROC LOSS FUNCTION IN PYTORCH

```
sofmx = nn.Softmax(dim=1)
def logistic_func(x,k, L=1, x_zero=0):
    return L/(1+ torch.exp(-k*(x-x_zero)))

class AUCLoss(nn.Module):
    def __init__(self):
        super(AUCLoss, self).__init__()

    def forward(self, output, target):
        output = sofmx(output)[:,1]
        pos_pred = output[ target == 1]
        neg_pred = output[ target == 0]
        pairwise_matrix = pos_pred.unsqueeze(1) - neg_pred
        transform = logistic_func(pairwise_matrix, 20)

        return 1-transform.mean()
```

ALGORITHM IV. IMPLEMENTATION OF MULTICLASS AUROC LOSS FUNCTION IN PYTORCH

```
class MulticlassAUCLoss(nn.Module):
    def __init__(self):
        super(MulticlassAUCLoss, self).__init__()

    def forward(self, output, target):
        output = sofmx(output)
        num_classes = output.size(1)
        sum_per_class_auc = 0.0

        for class_index in range(num_classes):
            pos_pred = output[:, class_index][target == class_index]
            neg_pred = output[:, class_index][target != class_index]
            pairwise_matrix = pos_pred.unsqueeze(1) - neg_pred.unsqueeze(0)
            transform = logistic_func(pairwise_matrix, 20)
            class_auc = transform.mean()
            sum_per_class_auc += class_auc

        return 1- (sum_per_class_auc / num_classes)
```

## IV. RESULTS AND DISCUSSION

PLGNT is the most common type of brain tumor in children, and CNNs have been shown to be effective for classifying BRAF fusion and BRAF V600E mutation PLGNT molecular subtypes on MR images. AUROC is a popular metric for evaluating CNNs, which is not differentiable. Thus, the models cannot be directly optimized for AUROC. In this study we introduced an AUROC loss function to improve the performance of PLGNT molecular subtype classifiers.

For a binary classifier whose output is a probability, ROC curve is formed by plotting true positive rates against false positive rates at different thresholds which are used to convert the output probability into a predicted label. AUROC is the area under ROC that is not differentiable. Inspired by WMW statistics, we developed an AUROC loss function in PyTorch, which is open-sourced in GenuineAI Python library.

We validated our loss function to improve the performance of CNN classifiers for PLGNT molecular subtype identification. We used a shallow CNN architecture and repeated the train-validation-test (60/20/20) experiment with different data splits and model random initializations 100 times. We achieved a test AUROC of 86.11 with 95% CI [84.96, 87.25] for BRAF fusion vs BRAF V600E mutation binary classification with BCE loss function. AUC loss improved the mean AUC to 87.71 CI [86.64, 88.79] (p-value 0.0456). In the context of multiclass classification, the AUROC improved from 74.42% with 95% CI [73.47, 75.38] to 76.59% with 95% CI [75.65, 77.53] (p-value 0.0016).

For the proposed pipeline the batch size was increased to 64, because AUROC is related to ranking and a higher batch size enables the model to rank the training examples more consistently. This is more important if the data is noisy and contains outliers. In the case that the computational resources do not allow increasing batch size, reducing learning rate and increasing the number of epochs with shuffling the train examples will help.

Directly optimizing CNNs for AUROC results in statistically significantly improved performance of CNN classifiers for PLGNT molecular subtype identification. Given that AUROC metric and the WMW loss function are closely related, increasing the batch size provides the CNN with a better context to optimize ranking of the examples during training.

AUROC is a well-known evaluation metric for binary classifiers which is frequently used in medicine. Our binary classification AUROC loss function improves the performance of different models such as deep PLGNT molecular subtype identifiers. Our implementation of multiclass AUROC is based on the one-versus-rest (OvR) method, which is more interpretable and less computationally expensive in comparison to one-versus-one (OvO).

It should be highlighted the proposed approach has limitations. The results show AUROC improvement through utilizing the loss function is statistically significant but marginal. As it was mentioned, a larger batch size is favorable, but may not be feasible due to memory limits. The minimum allowed batch size is 2, because at least one negative and one positive sample in each batch are required. Thus, if the batch size is low, the sampling should be controlled. Lastly, if the final goal is classification, using AUROC loss flips the challenge of training for classification (CE) and evaluating for ranking (AUROC) into training for ranking and evaluating for classification.

ACKNOWLEDGMENT

This research has been made possible with the financial support of the Canadian Institutes of Health Research (CIHR) (Funding Reference Number: 184015)